\documentclass[%
 reprint,
 amsmath,amssymb,
 aps,
pra,
]{revtex4-1}

\usepackage{graphicx}% Include figure files
\usepackage{dcolumn}% Align table columns on decimal point
\usepackage{bm}% bold math
\begin{document}

%\preprint{APS/123-QED}

\title{Intrinsic Optical Bistability in a Strongly-Driven Rydberg Ensemble }

\author{Natalia R. de Melo}\email{natalia.r.melo@durham.ac.uk}
\author{Christopher G. Wade}
\author{Nikola \v{S}ibali\'{c}}
\author{Jorge M. Kondo}
\author{Charles S. Adams}
\author{Kevin J. Weatherill}
\affiliation{Joint Quantum Centre (JQC) Durham-Newcastle, Department of Physics, Durham University, South Road, Durham DH1 3LE, UK}

\date{\today}% It is always \today, today,

\begin{abstract}

We observe and characterize intrinsic optical bistability in a dilute Rydberg vapor.  The bistability is characterized by sharp jumps between states of low and high Rydberg occupancy with jump up and down positions displaying hysteresis depending on the direction in which the control parameter is changed. We find that the shift in frequency of the jump point scales with the fourth power of the principal quantum number. Also, the width of the hysteresis window increases with increasing principal quantum number, before reaching a peak and then closing again. The experimental results are consistent with predictions from a simple theoretical model based on semiclassical Maxwell-Bloch equations including the effects of broadening and frequency shifts. These results provide insight to the dynamics of driven dissipative systems.

%\begin{description}
%\item[PACS numbers]check at the journal
%\end{description}
\end{abstract}

\maketitle

%\tableofcontents

\section{\label{sec:level1}Introduction\protect\ }

Optical bistability is a well studied phenomenon, having provided a rich contribution for the understanding of non-equilibrium systems~\cite{Haken1980, Marcuzzi2014}.  By definition, a system is bistable when, for the same input parameters, there are two stable output states. The classic conditions for observing optical bistability are a nonlinear system with feedback~\cite{Gibbs}.  Different kinds of system have been used to demonstrate bistability such as Fabry-Perot cavities~\cite{Gibbs}, nonlinear prisms~\cite{Stegeman1988}, photonical crystal cavities~\cite{wang2008}, QED cavities~\cite{kwon2013}, plasmonic nanostructures~\cite{wurtz2006} and nematic liquid crystals~\cite{Kravets2014}; and in general, for these examples, an optical cavity provides the feedback to the non-linear system.

Intrinsic optical bistability occurs when the bistable response is present in a system without an optical cavity and feedback is replaced by strong inter-particle interactions~\cite{Carmichael1977}. A dynamical equilibrium is reached between the driving of the excitation scheme and a dissipative process in the atom-light interaction. 
Recently, there has been much theoretical activity in exploring dynamics of these driven-dissipative systems \cite{Marcuzzi2014, Weimer2015, Nikola2015}. However, with observations to date limited to an up-conversion process using Yb$^{3+}$ ions in a solid-state crystal \cite{Hehlen1994} and more recently in strongly driven dilute Rydberg ensemble \cite{CCarr2013}, there is still a lack of experimental studies that would allow detailed analysis of observed bistabilities. 
\begin{figure}[b]
\includegraphics[width =8.5 cm,angle=0]{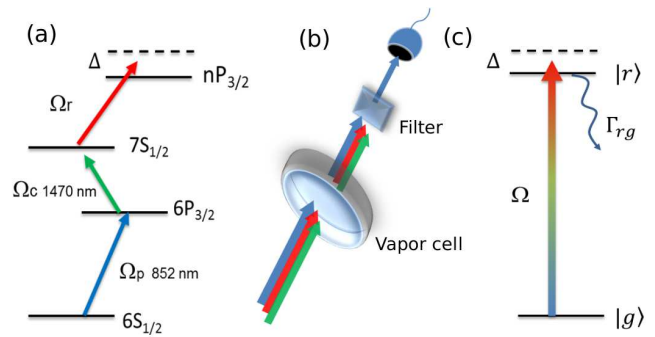}
\caption{\label{setup} (Color online) (a) Three-photon excitation scheme to Rydberg states in cesium. (b) Schematics of the experimental setup. The three excitation lasers copropagate through a 2 mm vapor cell. An interference filter is used to selected the transmission signal of the probe beam.(c) Simplified two-level model system.}
\end{figure}
In this work, we extend the experimental and theoretical study of the optical bistability in Rydberg ensemble \cite{CCarr2013} to investigate the behavior of optical bistability and hysteresis for a range of control parameters. By varying the principal quantum number $n$, of the Rydberg state, the atomic density and the excitation laser intensity we can control the driving and the dissipation within the system. Our experimental observations reveal a widening of the hysteresis window followed by a subsequent narrowing as $n$ is increased. We also observe a saturation of the bistability width dependent on the driving laser intensity. Furthermore, the results are consistent with predictions from a surprisingly simple theoretical model based on the semiclassical Maxwell-Bloch equations including the effect of level shifts and self-broadening originating from the interactions between Rydberg atoms. 
\section{Experiment}

The atomic excitation scheme and a schematic of the experimental setup are shown in figure~1(a) and (b). The Rydberg state is accessed via resonant three-photon driving~\cite{Carr12b} using a probe laser at 852 nm (Rabi frequency $\Omega_{{\rm p}}$) and a coupling laser at 1470 nm (Rabi frequency $\Omega_{{\rm c}}$) to drive the $\left|{\rm 6S}_{1/2}, F=4\right\rangle\rightarrow\left|{\rm 6P}_{3/2}, F'=5\right\rangle$ and $\left|{\rm 6P}_{3/2}, F'=5\right\rangle\rightarrow\left|{\rm 7S}_{1/2},F''=4\right\rangle$ transitions, respectively. The probe and coupling beam $1/e^{2}$ radius are 50 and 45 ${\rm \mu m}$, respectively, and the beam powers are typically set in the range 20 to 70 ${\rm \mu W}$. These lasers are stabilized on resonance using polarization spectroscopy~\cite{CCarr2012}.
The final `Rydberg' laser (Rabi frequency $\Omega_{{\rm r}}$) is scanned across the resonance between $7{\rm S}_{1/2}$ and a Rydberg state ${\rm nP}_{3/2}$. We access transitions to n${\rm P}_{3/2}$ states in the range $n =12$ to $50$, corresponding to a wavelength range from 882.70 to 780.03 nm, using a Ti:Sapphire laser. This laser has a $1/e^{2}$ radius of 45 ${\rm \mu m}$ and the typical power range is between 100 mW and 1 W. 
All lasers are copropagating and focused through the center of a 2~mm Cs vapor cell. Using an interference filter, we reject the coupling and `Rydberg' light and measure the transmission of the probe beam. By means of population shelving in the Rydberg level \cite{Thoumany2009} the transmission provides an effective readout of the population in the Rydberg state.
The atomic density in the sample is controlled by varying the temperature of the cell \cite{Gallagher73}, which is dynamically stabilized. 

\begin{figure}
\centering
\includegraphics[scale=0.95]{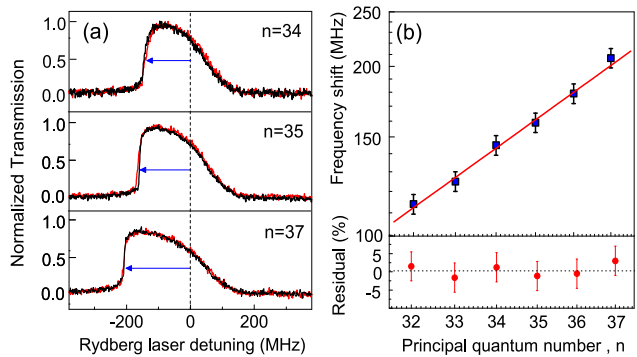}
\caption{(Color online) (a) Experimental optical response, the transmission of the probe beam as a function of the Rydberg laser detuning for the same Rabi frequency $\Omega_{{\rm r}}/2\pi= (80 \pm 10)$ MHz and different Rydberg levels $n=34, 35$ and $37$. $\Omega_{{\rm p}}/2\pi= (140 \pm 10)$ and $\Omega_{{\rm c}}/2\pi= (170 \pm 10)$ MHz (b) Measurement of the frequency shift of the phase transition as a function of the principal quantum number $\rm n$, of the Rydberg level $n$P$_{3/2}$. The red line is a linear fit showing that the shift scale with $\approx n^{4}$.}
\label{figure2}
\end{figure}

Figure 2(a) shows the transmission of the probe beam as a function of the Rydberg laser detuning $\Delta$, sweeping the frequency in positive (red) and negative (black) directions.  Each curve corresponds to the spectrum for a different nP$_{3/2}$ Rydberg state with $n=34$, $35$ and $37$. The results were obtained at atomic density of $(1.0\pm 0.5)\times 10^{10}$ cm$^{-3}$, with probe and coupling Rabi frequency, $\Omega_{{\rm p}}/2\pi= (140\pm 10)$  and $\Omega_{{\rm c}}/2\pi= (170 \pm 10)$~MHz. The Rabi frequency of the final step transition is kept constant at $\Omega_{{\rm r}}/2\pi= (80 \pm 10)\!$~MHz as $n$ is changed.
Under these conditions we observe an asymmetric lineshape with a sharp phase transition between the states of low and high Rydberg occupancy, shifted to the red side of resonance. 
 
The behavior of the measured shift as a function of $n$ is shown in figure 2(b). The frequency shift of the phase transition is seen to scale with the fourth power of the principal quantum number ($n^4$). This result is consistent with an interaction shift due the strong dipole-dipole interactions between Rydberg atoms, since the interaction potential between two dipoles is $V_{\rm dd} \propto \rm d^{2}$ where dipole moment d~$\propto n^{2}$. However, other interactions, such as ionizating collisions \cite{Olson1979} or dipole-dipole energy transfer \cite{Anderson1998} could have the same $n$ scaling.
\begin{figure}[b]
\includegraphics[scale=1]{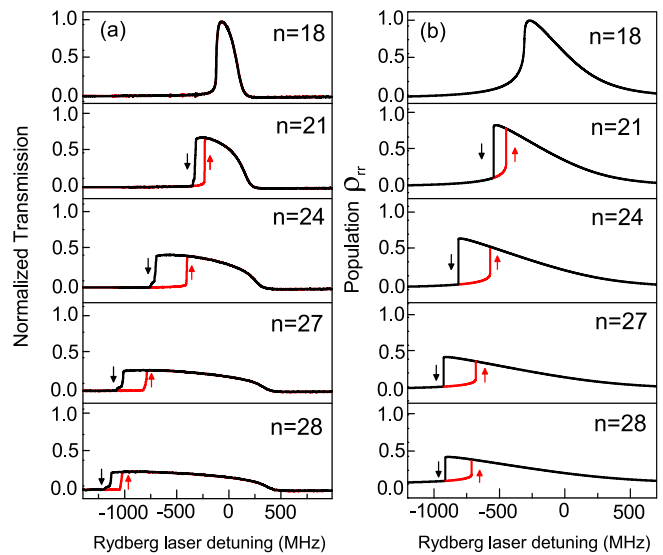}
\caption{\label{spectra_teta}(Color online) (a) Experimental optical response, the transmission of the probe beam as a function of the Rydberg laser detuning for the same Rabi frequency $\Omega_{{\rm r}}/2\pi=(130\pm 10) $ MHz and different Rydberg levels. (b) Theoretical Rydberg state population $\rho_{{\rm rr}}$ from the model as a function of the laser detuning for different Rydberg levels. Theoretical parameters: $\Omega_{\rm r}/2\pi=40$ MHz and $\Gamma_{\rm rg}/2\pi=3$ MHz.}
\end{figure}

Upon stronger driving or increased atomic density, above a critical limit, the behavior becomes more complex. The interaction-induced frequency shift surpasses the width of the optical resonance and results in a intrinsic optical bistability and hysteresis. Fig. 3(a) shows the transmission of the probe beam as a function of the Rydberg laser detuning for different Rydberg states, ranging from $n=18$ (top) to $n=28$ (bottom). This result is obtained at atomic density of $N=(3.0 \pm 0.5) \times 10^{11}$ cm$^{-3}$ and Rabi frequencies $\Omega_{{\rm p}}/2\pi=(130 \pm 10)$, $\Omega_{{\rm c}}/2\pi= (190 \pm 10)$ and $\Omega_{{\rm r}}/2\pi=(130 \pm 10 )\! $ MHz. Each curve shows positive (red) and negative (black) scan across the resonance for different Rydberg levels and are normalized to the peak transmission amplitude of the $n=18$ state. With the exception of the $n=18$ state, each scan displays obvious optical bistability with hysteresis dependent on the direction in which resonance is approached, indicated by arrows on the plots.
Within the hysteresis region the system can be maintained in two different steady states of high and low Rydberg occupancy for the same experimental parameters. As $n$ is increased the width of the hysteresis window initially becomes larger followed by a subsequent narrowing. The evolution of the bistability width with $n$ is shown in figure 4 for two different atomic densities $N=(2.0 \pm 0.5)\times10^{11}$ (purple squares) and  $(3.0\pm 0.5)\times10^{11}$ cm$^{-3}$ (green circles). The range of $n$ accessed is ultimately limited by the maximum power of the Rydberg laser in order to keep the Rabi frequency, $\Omega_{\rm r}$, constant. For low atomic density we observe a saturation of the bistability width for higher $n$, but for high atomic density we can observe a maximum value at $n=26$ and subsequent narrowing. 

\section{Theoretical model}

To understand this behavior we model the system using density-matrix formalism applied to a simple two-level system, as shown in Fig. 1(c). We consider a ground state $\left|g\right\rangle$ and a Rydberg state $\left|r\right\rangle$ coupled by a laser with Rabi frequency $\Omega$ and detuning $\Delta$, where $\Gamma_{\rm rg}$ is the decay rate from the Rydberg to the ground state. Using semi-classical analysis, the time evolution of the system is described by a Lindblad master equation applied for a simple two-level system \cite{Lindblad1976}. 
Typically mean-field theory is used to describe the dipole-dipole interaction between the Rydberg atoms in a classical approximation \cite{lee2012}, where a many-body interaction system is described in terms of the response of a single atom interacting with a mean-field interaction potential. This results in renormalization of the transition frequency $\Delta \rightarrow \Delta_{{\rm eff}}=\Delta-\Delta_{{\rm dd}}$, where $\Delta_{{\rm eff}}$ is an effective detuning \cite{lee2012} and $\Delta_{{\rm dd}}$ is the mean-field shift which can be expressed as $\Delta_{{\rm dd}}=V \times \rho_{{\rm rr}}$, where $\rho_{{\rm rr}}$ is the fraction of atomic population in the Rydberg state and $V$ is the interaction term corresponding to the sum of the dipole-dipole interaction over the excitation volume. Under this condition, the optical Bloch equations (OBE) are modified and the time evolution of the matrix elements are:
 \begin{subequations}
  \begin{align}
&\dot{\rho_{{\rm gr}}}={\rm i} \Omega \left(\rho_{{\rm rr}}-1/2\right)+{\rm i}\left(\Delta_{{\rm eff}}-{\rm V}\rho_{{\rm rr}}\right)\rho_{{\rm gr}}-\frac{\Gamma_{\rm rg}}{2}\rho_{{\rm gr}} \\
&\dot{\rho_{{\rm rr}}}=-\Omega {\rm Im}(\rho_{{\rm gr}})-\Gamma_{\rm rg} \rho_{{\rm rr}},
\end{align}
\end{subequations}
where the coherence terms $\rho_{{\rm gr}}=\rho_{{\rm rg}}^{*}$ and the diagonal population terms are related as $\rho_{{\rm gg}}=1-\rho_{{\rm rr}}$. $\rho_{{\rm gg}}$ and $\rho_{{\rm rr}}$ are the fraction of the atomic population in the ground and Rydberg state, respectively. 
The steady-state solution can be obtained assuming $\dot{\rho}_{{\rm gr}}=\dot{\rho}_{{\rm rr}}=0$ and solving the set of equations (1).  

Substituting for $\Delta_{{\rm eff}}$ in the steady-state solution of Eq~1, we obtain a cubic equation for the population in the Rydberg state:
\begin{eqnarray}
\frac{\Omega^{2}}{4}-\left(\frac{\Omega^{2}}{2}+\frac{\Gamma_{\rm rg}^{2}}{4}+\Delta^{2}\right)\rho_{{\rm rr}}-2\Delta V \rho_{{\rm rr}}^{2}-V^{2}\rho_{{\rm rr}}^{3}=0.\ \ \ 
\end{eqnarray}
This equation provides the steady-state solution for the Rydberg population as a function of the Rydberg laser detuning. Within the frequency range of the hysteresis window there are three  distinct solutions,  one unstable and two stable that correspond to the two steady-states of the bistable regime.   
This basic model can explain the origin of the optical bistability in the experimental system as shown in reference \cite{CCarr2013} but does not capture features of the experimental data in figure 3(a), such as the closing of the bistability width at high density.

We therefore extend our model to include the effect of a self-broadening contribution where the total relaxation rate is rewritten as:
\begin{eqnarray}
\Gamma_{\rm rg}\rightarrow \Gamma_{\rm rg}+\Gamma_{{\rm self}}=\Gamma_{\rm rg}+\beta N~, 
\end{eqnarray}
where $\beta$ is the self-broadening coefficient and $N$ is the atomic density.  

The self-broadening effect arises from the interaction between two identical atoms in a superposition of the ground and excited states, so that all dipole-allowed transitions could contribute to the effect. Indeed, in our system, different mechanisms may contribute for the broadening effect, like associative ionization \cite{Chret1982} or photoionization \cite{galagher}, also different collisional process, as state-mixing, for instance \cite{Stoicheff1980, Reinhard2008}, but we are not able to distinguish each individual contribution. 
Self-broadening coefficients have been studied in atomic vapors for many years particularly for the D2 lines of alkali-metal atoms~\cite{lewi1980, weller2011}. However, in this work on Rydberg states, we expect the broadening to have a number of difference contributions. For simplicity, we apply the same $n^4$ scaling observed for the frequency shift also to the self-broadening term
and write the self-broadening coefficient as: 
\begin{eqnarray}
\Gamma_{{\rm self}}=\beta' n^{*4} N ,
\end{eqnarray}
where $n^{*}$ is the effective principal quantum number of the Rydberg state $n$P$_{3/2}$.

\section{Analysis and Discussion}

In figure 3(b) we present the results of our model and plot Rydberg state population $\rho_{{\rm rr}}$ as a function of the laser detuning, for several Rydberg levels. To compare these results with the experimental data plotted in figure 3(a), the curves are normalized to the maximum transmission amplitude of the $n=18$ scan. For this calculation, the theoretical parameters are $\Omega/2\pi= 40$ MHz and $\Gamma_{{\rm rg}}/2\pi=3$ MHz. By applying an n$^4$ scaling, the parameter V is found by fitting the model to  the experimental results and is write as $V=8.9 \times 10^{-14} \times$n$^{*4}N$ MHz cm$^{3}$. 

The results shows good qualitative agreement between experiment and theory with respect to the amplitude of the transmission signal and the width of the hysteresis window. The addition of the self-broadening-like term to the model is essential to reproduce the experimental trends observed at higher density and stronger driving.

In figure 4, a quantitative comparison is shown. This result shows the bistability width as a function of the principal quantum number for two atomic densities, $N=(2.0\pm 0.5)$ and $(3.0\pm 0.5) \times 10^{11}$ cm$^{-3}$. In this picture, the full circles and square with the error bars are the experimental data, and the column bars are the theoretical results considering the self-broadening effect. With the fitting of the figure 4 at low  density (purple squares) we find that the ratio $\Omega/\Omega_{\rm r}\approx 1$ which is expected as the transition of the two first excitation steps are saturated.  
\begin{figure}[b]
\includegraphics[width =6.3cm,angle=0]{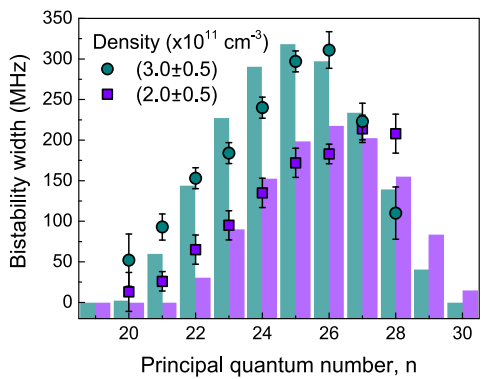}% Here is how to import EPS art
\caption{(Color online) Bistability width as a function of the principal quantum number for two atomic densities, $N=(2.0\pm 0.5)$(purple square) and $(3.0\pm 0.5)\times 10^{11} \: {\rm cm}^{-3}$(green circles), for the same Rydberg Rabi frequency, $\Omega_{\rm r}/2\pi=(120 \pm 10)$ MHz. The column bars are the theoretical results considering the self-broadening effect, $\beta'/2\pi=(2.0\pm 0.2) \times 10^{-9}$ Hz cm$^{3}$ and the parameters $\Gamma_{\rm rg}/2\pi=13.5$ MHz and $\Omega/2\pi=(140\pm 5)$ MHz.}
\end{figure}
Also fitting the model to the data of figure 4, we get a value for the broadening coefficient, $\beta'/2\pi=(2.0\pm 0.2) \times 10^{-9}$ Hz cm$^{3}$, which is a key quantitative result to characterize the system. Comparing this result with the literature \cite{haroche1981}, we find that $\beta'{\rm n}^{4}$ is about an order of magnitude larger than the D$_{2}$ line self-broadening coefficient in cesium \cite{weller2011} and results fouond in line-broadening experiments using Doppler-free spectroscopy \cite{Thompson1987}. However these experiments were performed with lower atomic states in contrast to our Rydberg states where the strong interactions and multiple contributions to the broadening term is a distinguishing characteristic \cite{galagher}. Our result is in agreement with an  experimental result for the self-broadening coefficient involving Rydberg states, achieved in four-wave mixing in a Rb vapor \cite{Melo2014}.

Finally, we characterize the width of the bistability window for a fixed Rydberg state ($n=28$) as a function of Rydberg laser power for three different atomic densities shown in figure 5: $N=\left\{3.0 (\rm O) ; 1.0 (\square) ; 0.7 (\triangle)\right\} \pm 0.5 \;\times 10^{11}$ cm$^{-3}$.
The solid lines represent the theoretical calculation for the each density. The Rabi frequency is related to the Rydberg laser power $P$ by $\Omega\propto\sqrt{P}$. By again finding the best fit of the model to the data, we find $\Omega/2\pi=(137 \pm 2) \times \sqrt{P}$ which confirms the ratio $\Omega/\Omega_{\rm r}\approx 1$ as $\Omega_{\rm r}/2\pi=(130 \pm 10) \times \sqrt{P}$ for the experimental data.  Also, the resultant value of the self-broadening coefficient, $\beta'/2\pi= (1.6 \pm 0.2) \times 10^{-9}$ Hz cm$^{3}$, is consistent with the value found in the previous fitting (figure 4).
\begin{figure} [t]
\includegraphics[width =6.3 cm,angle=0]{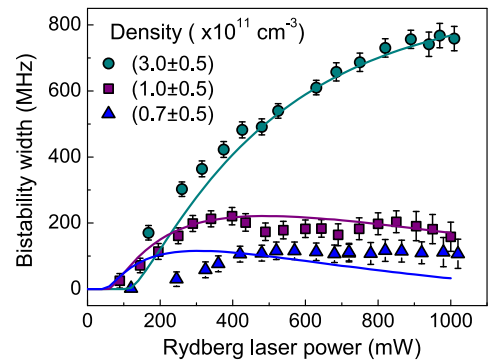}% Here is how to import EPS art
\caption{(Color online) Bistability width as a function of the Rydberg laser power for n$=28$ and three different atomic densities $N=\left\{3.0 (\rm O) ; 1.0 (\square) ; 0.7 (\triangle)\right\} \pm 0.5 \;\times 10^{11}$ cm$^{-3}$. The solid lines are the theoretical calculation using the parameters: $\beta'/2\pi= (1.6 \pm 0.2) \times 10^{-9}$ Hz cm$^{3}$, $\Gamma_{{\rm rg}}/2\pi=13.5$ MHz and $\Omega/2\pi=(137 \pm 2) \times \sqrt{P}$,  where $P$ is the Rydberg laser power.}
\end{figure}\vspace{-0.5 mm}
Although the theoretical model doesn't reproduce the oscillatory feature present in the data for N$=1.0\pm 0.5 \;\times 10^{11}$ cm$^{-3}$, this result shows good agreement especially for high density and we are able to describe the mean behavior of the bistability width.
  
\section{Conclusions}

In conclusion, we have investigated intrinsic optical bistability in cesium vapor, in a multi-photon excitation scheme to Rydberg states. We characterized the behavior of the bistability width as a function of the principal quantum number and the Rydberg laser power for Rydberg states $n$P$_{3/2}$ with $n$ between 18 and 37, for different atomic densities. We observed that the shift of the phase transition scales as $n^4$ and the width of the hysteresis window exhibits a maximum value with a subsequent narrowing with the increasing of the principal quantum number and a saturation behavior as a function of the Rydberg laser power.  A theoretical model using the OBE for a simple two level system, with modification to include effects of frequency shift and broadening, reproduce the behaviour of our experimental observation and allows us to estimate the self-broadening coefficient for the studied Rydberg levels. 
We believe that the observations reported are important for ongoing studies in Rydberg non-linear optics \cite{Prit13} especially those employing multi-step excitation schemes~\cite{Kond15}. The work provides the first detailed characterization of system displaying intrinsic optical bistablity and yields insight for theoretical descriptions of such systems. The data
presented in this paper are available \cite{data}.
\begin{acknowledgments}
This work was supported by Durham University, The Federal Brazilian Agency of
Research (CNPq), and EPSRC ”(Grant No.EP/M014398/1 and EP/M013103/1)”.
\end{acknowledgments}

%\nocite{*}


\end{document}